# STRETCHING VIBRATIONS OF C-Br IN P-DIBROMOBENZENE NANOPARTICLES

**M.A.Korshunov**
*Institute of physics it. L. V.Kirenskogo of the Siberian branch of the Russian Academy of Sciences, 660036 Krasnoyarsk, Russia*
*e-mail: mkor@iph.krasn.ru*

Raman spectrums of nanoparticles of a p-dibromobenzene with size of 300 nanometers and 50 nanometers are measured. It is found that frequencies of lines of stretching vibrations of C-Br at reduction of the sizes of nanoparticles below 300 nanometers are incremented. It is related to magnification of parameters of the lattice and interaction reduction between molecules.

By method of the molecular dynamics the formation of nanoparticles of an organic molecular crystal of a p-dibromobenzene was investigated. It is shown that if one considers interaction not only of the nearest molecules, with the increase of particle sizes the lattice parameters decrease. It, apparently, should affect not only the lattice vibrations but also the intramolecular ones and should appear in Raman spectrums. At magnification of parameters of a lattice, the interaction between molecules decrease and frequencies of stretching vibrations of C-Br should be incremented and come nearer to frequencies of molecules which are observed in a p-dibromobenzene in liquid or a gaseous state. In the vibration spectrum of a solution of a p-dibromobenzene, the value of frequency of stretching vibrations C-Br $\nu=215$ cm$^{-1}$ [1], and in a volume single crystal it is $\nu=212$ cm$^{-1}$.

Nanoparticles of a p-dibromobenzene with the size of 300 nanometers and 50 nanometers have been synthesized. The size of nanoparticles was measured by an electronic microscope. After particle-size determination, the measurements of Raman spectrums on spectrometer Jobin Yvon T64000 was made. In Fig. 1 spectrums of intramolecular oscillations for particles with a size 300 nanometers (1) and 50 nanometers (2) are given.

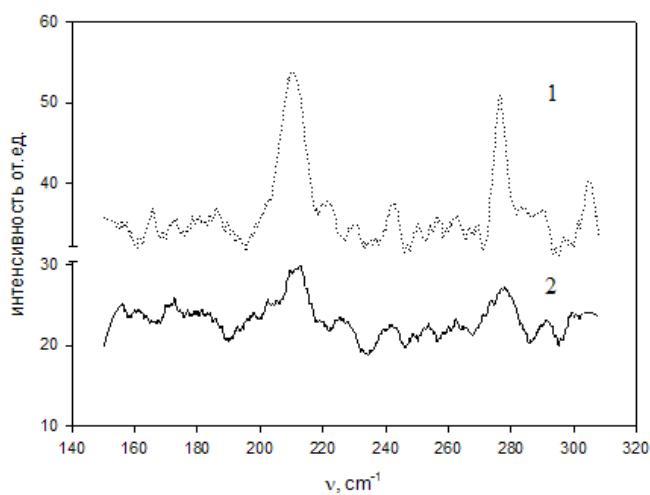

*Fig. 1. A spectrum of intramolecular oscillations of nanoparticles of a p-dibromobenzene with the size of 300 nanometers (1) and 50 nanometers (2) in range from 150 to 310 cm$^{-1}$.*

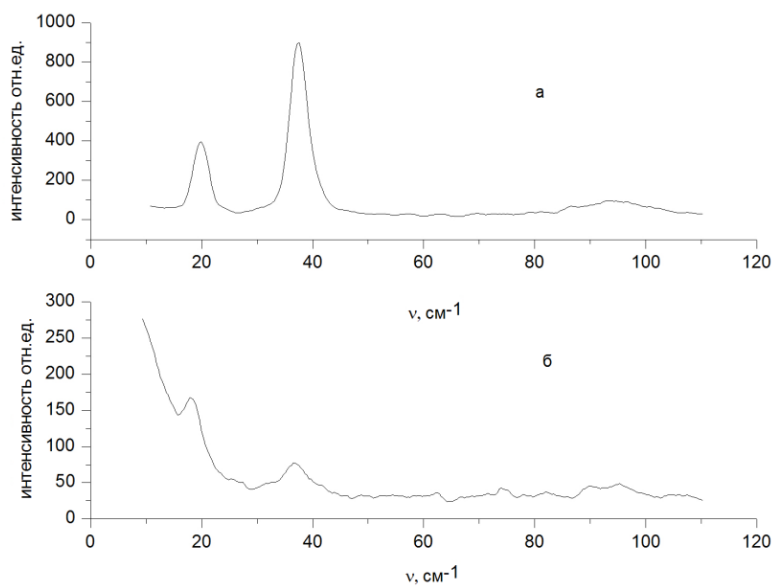

*Fig. 2. A spectrum of the lattice oscillations of a p-dibromobenzene for the size of nanoparticles 21μm and 50 nanometers.*

Value of frequency of a spectral line of a stretching vibration of C-Br for the size of a nanoparticle of a p-dibromobenzene of 300 nanometers is equal to 212 cm$^{-1}$, and for a grain size of 50 nanometers it is 214 cm$^{-1}$. At the same time, values of frequencies of the lattice vibrations are lowered (Fig. 2). Measurements show that at change of the sizes of crystals from 300 μm to ~300 nanometers values of frequencies of studied intramolecular vibrations are not changed.

Lowering of frequencies of the lattice vibrations at reduction of the sizes of nanocrystals is caused by magnification of parametres of a lattice. The magnification of parametres affects intermolecular interaction reduction. It, apparently, promotes increase of intramolecular vibrations of C-Br that observed in spectrums.

Thus, at reduction of the sizes of nanoparticles of a p-dibromobenzene below 300 nanometers up to 50 nanometers change of intramolecular vibrations (in particular stretching vibrations) of C-Br from 212 cm$^{-1}$ to 214 cm$^{-1}$ is observed. It can be caused by the magnification of parameters of a lattice, which is confirmed by decrease of frequencies of the lattice vibrations.